\documentclass{aa}
\usepackage{graphics}
\begin{document}

\thesaurus{11(11.13.2;11.19.6)}

\title{A simple model of magnetically induced warps}

\author{E. Battaner \and J. Jim\'enez-Vicente}

\institute{Depto. de F\'{\i}sica Te\'orica y del Cosmos. Univ. de Granada.\\
18071 Granada.
Spain}

\maketitle
\markboth{E. Battaner \& J. Jim\'enez-Vicente:
A simple model of magnetically induced
warps}{A simple model of magnetically induced
warps}

\begin{abstract}

        Assuming the magnetic hypothesis for the formation of warps, we deduce
a very simple formula for the warp curve in an idealized scenario. 
According to this formula the warp rises as the third power of radius in the
innermost warped region, 
reaches the maximum slope at intermediate radii and has an asymptotic
slope coincident with the direction of the extragalactic magnetic field. In
most cases, however, the galaxy's limited size prevents the observation of
the full curve. Even though the model is highly simplified, it
basically reproduces real warp curves, in particular the 21 cm warp
curve of NGC 5907. If the magnetic model were considered to be
correct, the fitting of warp curves could allow rough estimations of
the strength and direction of the magnetic field. We also propose a
magnetic field distribution for the outermost part of the galaxy,
matching the boundary conditions of being azimuthal inside, and
constant at infinity. We use this magnetic field distribution to show
that the assumptions made to obtain the warp curve with our simple
model cannot introduce important errors.

\end{abstract}

\section{Introduction}

        The possibility of warps being induced by extragalactic magnetic 
fields (Battaner et al., 1990) remains a tempting explanation of this common
feature of most spiral galaxies. Without intending to enter a discussion about
its validity against other current models, some observational
facts may produce a renewed interest in the magnetic model: a) Binney (1991) 
and Combes (1994) argued that the required magnetic fields were too high for the
intergalactic medium. However, Kronberg (1994) reviewed evidence for magnetic
fields of about 1-3$\mu {\mbox G}$ to be common in intra-cluster
media.
Later, Feretti et al. (1995) have found $B < 8.3 \mu {\mbox G}$ in the
Coma cluster, about one order of magnitude higher than the
equipartition value.
This is a critical point, as magnetically driven warps require magnetic
strengths of this order of magnitude. Further confirmation of high intergalactic
fields would support our hypothesis. We are not aware of any published
alternative interpretation of the available observations of intergalactic magnetic
fields.
b) One of the most
interesting possibilities is that of warps being normal modes of oscillation
of the galactic disk (Sparke and Cassertano, 1988). But when the back reaction
of the halo is taken into account 
this model seems to fail (Dubinski and Kuijken, 1995; Kuijken, 1997). 
c) Zurita and Battaner (1997) have shown the 
tight alignment of the warps of the three largest spirals in the Local Group.
Such an alignment is easily understood in the magnetic model, and probably also
in the model by Kahn and Woltjer (1959).
        
        In this short paper we obtain a simple formula describing the warp 
curve [mean $z$ versus $x$] when the magnetic hypothesis is adopted. 
The coordinate $x$ is contained in the inner unwarped disc plane and is
perpendicular to the line of nodes, assumed to be untwisted for simplicity. 
Very simple models
for idealized scenarios are specially appropriate for exploratory hypotheses
and remain useful even when more sophisticated computations are available.

\section{The shape of magnetically induced warps}

        According to Battaner et al. (1990), the vertical magnetic force which induces 
the warp is of the order $B_y B_z/8 \pi L$, where $B_y$ is the component of the
extragalactic field in the equatorial plane of the galaxy perpendicular to
the line of nodes, $B_z$ is the component in the direction of the rotation
axis of the galaxy and $L$ is an undetermined characteristic
length. To deduce this expression, we assume a constant mean value of
the magnetic field strength in a large characteristic length
connecting the inner disc and the intergalactic medium. This seems to
be an extremely simple assumption, but it will be shown later that it
does not introduce a quantitatively important error.
Let us assume in this simplified model that these quantities $B_y$,
$B_z$, and $L$ are constants in the region of interest. 

The
characteristic length $L$ would depend on the degree of ionization
which in turn depends on the galactocentric radius. Let us however
assume that at any radius the degree of ionization is large enough as
to assure infinite conductivity and frozen-in magnetic field lines.
In the inner part of the disc this force is
negligible compared with the gravitational force, but it becomes increasingly
important towards the outer parts of the disc.

\begin{figure*}
\resizebox{18cm}{!}{\includegraphics{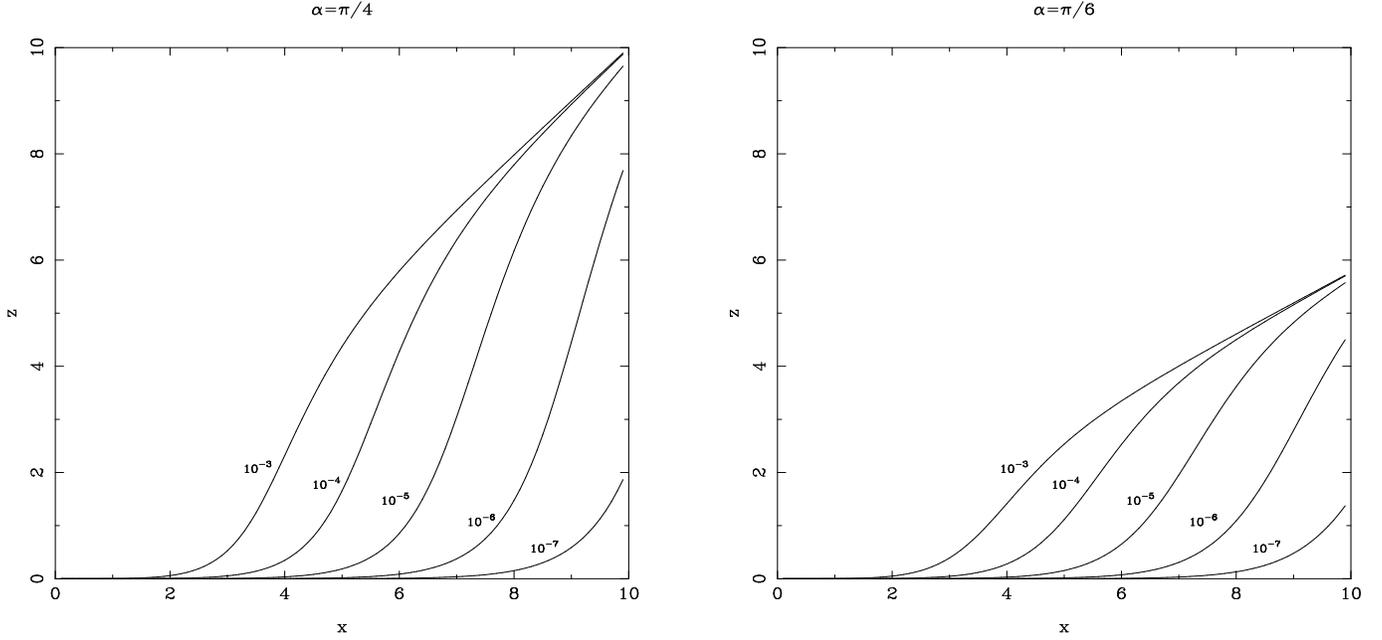}}
\caption[]{Warp curves for magnetically induced warps. Values of $k$ 
are indicated beside each curve and the value of $\alpha$ is 
indicated at the top of the plot
\label{fig1}}
\end{figure*}

        The expression is valid for a point in the equatorial plane of the
galaxy. For a point above this plane the vertical force becomes
\begin{equation}
F_z=\frac{B^2 \sin(2(\alpha - \beta))}{16 \pi L} \cos{\beta}
\end{equation} 
where $\tan \beta =z/x$ and $\tan \alpha$ is the slope of the direction of the
magnetic field in a $[x,z]$ diagram. It is easily derived that this expression
is equivalent to
\begin{equation}
F_z=\frac{B^2}{16 \pi} \frac{x(x^2-z^2)}{r^3}\sin (2\alpha) - 
\frac{2zx^2}{r^3}\cos (2\alpha)
\end{equation}
where $r=(x^2+z^2)^{1/2}$ is the radial coordinate in the $[x,z]$ plane

        In this simple model we assume that the gravitational potential is that
of a point mass in the centre of the galaxy, as a simplifying assumption
for the outer part of a disk not embedded in a massive halo. This
potential has already been used before, for instance by Cuddeford and Binney
(1993). Equilibrium in the vertical direction gives
\begin{eqnarray}
\frac{\partial p}{\partial z}& + & \rho GMr^{-3}z= \nonumber \\
 & & \frac{B^2}{16 \pi L} \left( 
\frac{x(x^2-z^2)}{r^3}\sin (2\alpha) - \frac{2zx^2}{r^3}\cos (2\alpha) \right)
\end{eqnarray}

        A full solution for the distribution of the gas in the
combined magnetic-gravity force field would reproduce the whole
geometry of the warp. This full solution is beyond the scope of this
paper. In any case, it would be interesting to, simply bu precisely
define the warp curve. In the inner unwarped region, 
$\frac{\partial{p}}{\partial{z}}=0$ at $z=0$, in the galactic plane. Let
us therefore define the warp curve as the locus of points where $\partial p/ 
\partial z =0$. We adopt an exponential law for the
disc density with length scale $R$, i.e. $\rho =\rho_0 e^{-x/R}$. Using
$R$ as length unit for $x$ and $z$, we obtain for the warp curve
\begin{eqnarray}
z & = & [ -e^{-x}-2kx^2\cos (2\alpha)+\{(e^{-x}+2kx^2 \cos (2\alpha))^2
\nonumber \\
 & & +4k^2x^4\sin (2\alpha)\}^{1/2} ] (2kx\sin (2\alpha))^{-1}
\end{eqnarray}
where
\begin{equation}
k=\frac{R^2 B^2}{16 \pi L \rho_0 GM}
\end{equation}
is one of the adjustable parameters which compares the extragalactic magnetic
energy density with the gravitational energy density. The other adjustable
parameter is $\alpha$ which specifies the direction of the extragalactic
magnetic field. The warp curve is therefore defined with just two free
parameters: $k$ and $\alpha$.

        For small values of $x$ a series expansion gives
\begin{equation}
z=k\sin (2\alpha) x^3
\end{equation}
which is a very simple expression for small warps. This simple formula
illustrates the fact that the maximum efficiency in producing warps is obtained
for $\alpha=45^\circ$, as in Battaner et al. (1990)

        For very large values of $x$, we obtain
\begin{equation}
z=\tan (\alpha ) x
\end{equation}
i.e. the slope of the warp curve matches the direction of the magnetic field
at large radii.

        In Fig. \ref{fig1} we plot the obtained curves for
$k=10^{-3},10^{-4}$ $,10^{-5}, 10^{-6}, 10^{-7}$ for $\alpha =\pi /4$ 
and $\alpha = \pi /6$. The curves for $\alpha =\pi /4$ seem to be unrealistic
at first glance. Note, however, that in practice values at $x>>6$ are
unobservable (or the galaxy simply does not exist at these radii). For instance
if the $[\alpha =\pi /4, k=10^{-5}]$ curve is truncated at $x\approx 6$,
the obtained warp curve becomes quite familiar. We reproduce the curve for
larger values of $x$ and $z$ in order to see the region where the slope becomes
equal to the direction of the extragalactic magnetic field, which probably takes place
for radii far from observational capacities or galaxy limits.

        It is worth noting that, for $\alpha \leq \pi /4$, there is a 
change in slope. It is higher at intermediate regions before reaching its
asymptotic value $\tan \alpha$. This is a common feature of real warps, and
can even be directly observed in the early contour maps of NGC 5907 and NGC 4565 by
Sancisi (1976).

\begin{figure}
\resizebox{8.8cm}{!}{\includegraphics{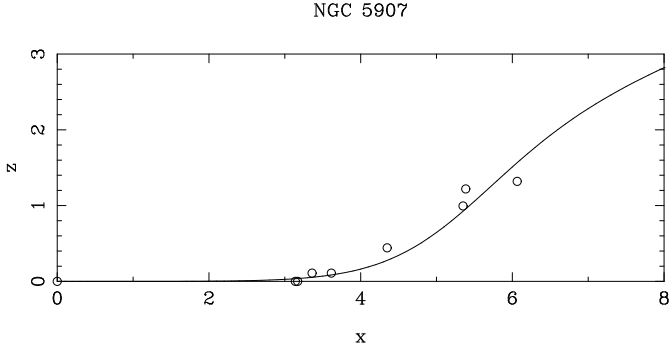}}
\caption[]{Warp curve for NGC 5907. Open symbols are experimental data adopted
from Sancisi (1976) and the solid line corresponds to the model
for $k=8\times 10^{-5}$ and $\alpha =20^\circ$. 
\label{fig2}}
\end{figure}

        Figure \ref{fig2} reproduces the observational curve for one of the
best known prominent warps, in NGC 5907,
adopted from Sancisi (1976). We cannot exclude that the warp of this
galaxy is due to other mechanisms, but we choose this warp because it
is one of the most representative and is studied very often. This
figure also reproduces a fitting to the model, with 
parameters $k=8\times 10^{-5}$ and $\alpha =20^\circ$. 
It is not straightforward to deduce the magnetic field strength from
the value of $k$, mainly because $L$ is an equivalent undetermined quantity.
For $L=1~\mbox {kpc}$, $\rho_0=1.7 \times 10^{-24}~\mbox{g cm}^{-3}$, $M=2\times
10^{11}~M_{\sun}$ 
and $R=4~\mbox{kpc}$, it is obtained that values of $k$ in the range $10^{-4}$
to $10^{-6}$ correspond to field strengths between $3~\mu \mbox{G}$ 
and $0.3~\mu \mbox{G}$,
in agreement with reported measurements by Kronberg (1994). Probably 
better results would probably be obtained with more realistic
calculations, but given the present, still exploratory, character of the 
magnetic model it is preferable to deal with idealized systems. The
noticeable fitting of the NGC 5907 warp curve suggest that the
magnetic model and the approximations considered here are not unreasonable

\section{Magnetic field distribution in the outer disc}

We have studied some magnetic effects in the density distribution of
the outermost disc without considering in detail the magnetic field
distribution in this region. This was made on one hand  because this
distribution is unknown, and on the other hand to produce an
exploratory simplified model. Now, we propose a magnetic field
distribution and use it to estimate the error of our assumption for
the specific case of $\alpha=45^\circ$, small warps and large galactocentric
radii.

As no radio continuum measurements are available to deduce the
magnetic field in this region, we must just assume it. For this task
we have not so many degrees of freedom, because our field must fulfill
two restrictions: a) Boundary conditions: it should be azimuthal
($B_{\varphi}$) at the inner radius $L_1$, and should become constant at
infinity. b) The field must also have vanishing divergence. We thus
solve the general problem of connecting both, the disc magnetic field
and the extragalactic magnetic field in a way that could be adopted in
other problems not associated with the particular case of warps. Outside the
galaxy the field could have any direction. To consider a general case
we take $B_x=B_z=B_{\infty}/ \sqrt{2}=A$ outside, but $B_y=0$, due to a proper
choice of the $y$ axis. Let us define: 

  \begin{equation}
    B_{1y}=-A \frac{xy}{\sigma^2} e^{-\frac{x^2+y^2}{2 \sigma^2}}
    \label{biy}
  \end{equation}
  \begin{equation}
    B_{1x}=\frac{A}{\sigma^2}e^{-\frac{x^2+y^2}{2
    \sigma^2}}(y^2-\sigma^2)+A
   \label{bix}
  \end{equation}
  \begin{equation}
    \label{bgx}
    B_{2x}=B_{\varphi} \frac{y}{|x|} \left[ 1 - \left(\frac{y}{x}\right)^2
    \right]^{-1/2} \frac{L_2-(x^2+y^2)^{1/2}}{L_2-L_1}
  \end{equation}
  \begin{equation}
    \label{bgy}
    B_{2y}=-B_{\varphi} \frac{x}{|x|} \left[ 1 - \left(\frac{y}{x}\right)^2
    \right]^{-1/2} \frac{L_2-(x^2+y^2)^{1/2}}{L_2-L_1}
  \end{equation}

         We then consider three regions:
\begin{enumerate}
\item Inner disc, for $0<r<L_1$. We do not propose any field
  distribution for this region, as our model only considers the
  periphery of the galaxy.
\item Outer disc, for $L_1<r<L_2$, taking $L_2$ as the radius where
  the galaxy's edge is assumed to be. In this region,
\begin{equation}
B_x=B_{1x}+B_{2x}
\end{equation}
\begin{equation}
  B_y=B_{1y}+B_{2y}
\end{equation}
\item Nearby intergalactic space, for $r>L_2$. We propose in this
  region,
  \begin{equation}
    B_x=B_{1x}
  \end{equation}
  \begin{equation}
    B_y=B_{1y}
  \end{equation}
\end{enumerate}

\begin{figure}
\resizebox{8.8cm}{!}{\includegraphics{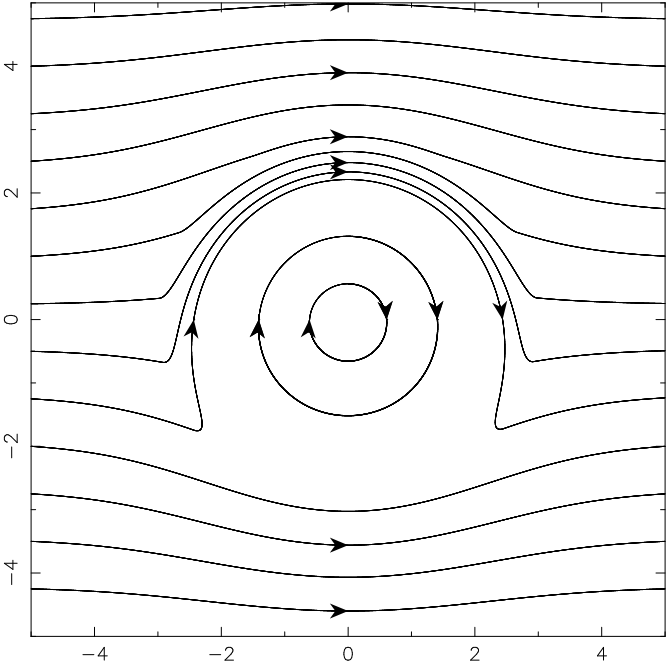}}
\caption[]{Magnetic field distribution of the galactic disc in the XY
  plane for
  $B_\varphi=3\mu G$, $A=0.2 \mu G$, $L_1=2$, $L_2=3$ and $\sigma=2$
  (taking the radial scale length as unit). 
\label{fig3}}
\end{figure}
\begin{figure}
\resizebox{8.8cm}{!}{\includegraphics{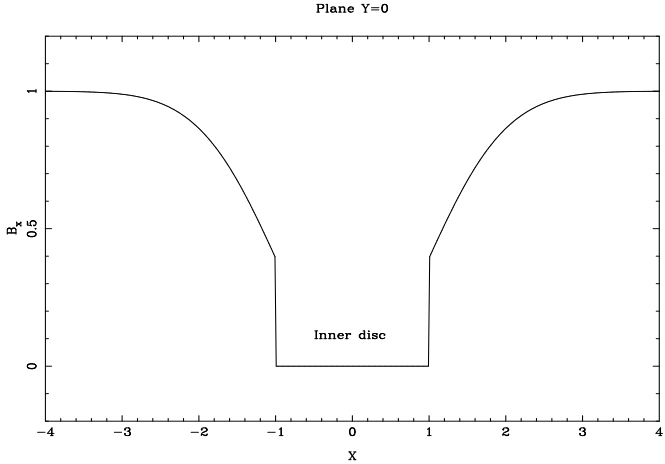}}
\caption[]{Magnetic field distribution of the galactic disc in the Y=0
  plane for
  $B_\varphi=3\mu G$, $A=0.2 \mu G$, $L_1=1$, $L_2=3$ and $\sigma=1$.
\label{fig4}}
\end{figure}

It can be easily checked that the divergence of this two-dimensional
field is zero. In Fig. \ref{fig3}, we plot this magnetic field
distribution. The magnetic field strength along $x$ is plotted in
Fig. \ref{fig4}. This distribution has no jump (only the first
derivative
is not a continuous function). In Fig. \ref{fig4} there is an apparent
jump at $L_1$, but it only means that we are ot interested in the
inner disc.

 Some of the formulae above would become clearer in polar coordinates,
 but cartesian coordinates are the natural ones outside the galaxy,
 and a choice covering the whole region is preferable.

We must complete our choice by defining the $z$ component. We assume
that $B_z$ is independent of $z$, in order to conserve the vanishing
divergence. Then, for $L_1<r<L_2$
\begin{equation}
  \label{bz}
  B_z=A\frac{(x^2+y^2)^{1/2}-L_1}{L_2-L_1}
\end{equation}
For $r>L_2$ we adopt $B_z=A$.

In this case, in the direction of the warp plane (perpendicular to the
line of nodes and to the galactic plane)
\begin{equation}
  \label{pbzx}
  \frac{\partial B_z}{\partial x}(y=0)=\frac{A}{L_2}
\end{equation}
as in Battaner, Florido and S\'anchez-Saavedra (1990).
\begin{figure}
\resizebox{8.8cm}{!}{\includegraphics{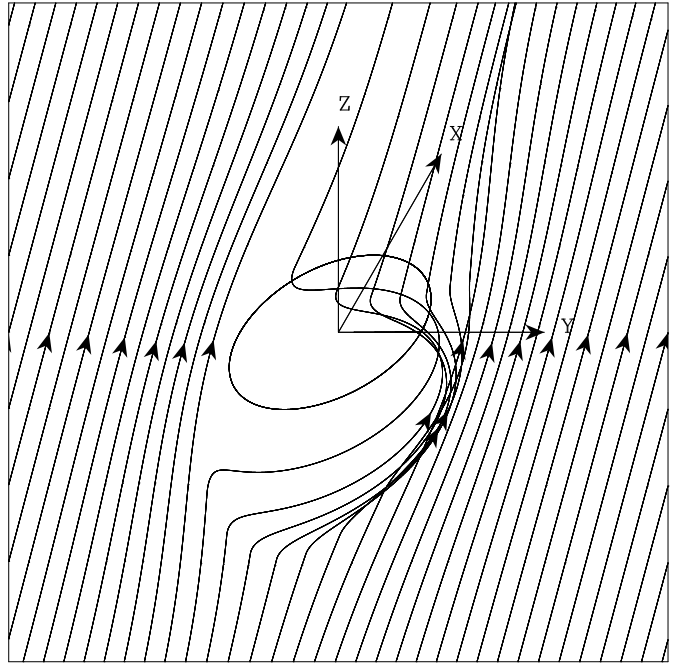}}
\caption[]{Magnetic field lines in the outer disc and in the
  intergalactic space close to a spiral galaxy. The warp plane is the
  $y=0$ plane. 
\label{fig5}}
\end{figure}
The three-dimensional picture of the magnetic field distribution can
be appreciated by combining figure (\ref{fig3}) and
eq. (\ref{bz}). The result is depicted in figure (\ref{fig5}), where
the inner azimuthal field lines have not been plotted.

Let us now consider a second model, with the simplifying assumption of
$\alpha=45^\circ$ and $z_{warp} \ll x_{warp}$, but taking into account
the above proposed magnetic field configuration.

The vertical force will be:
\begin{eqnarray}
  \label{vert}
  F_z & = & \frac{1}{4\pi}B_x \frac{\partial B_z}{\partial x}
  \nonumber \\
      & = & \frac{1}{4\pi} ( \frac{A}{\sigma^2}e^{-\frac{x^2+y^2}{2
    \sigma^2}}(y^2-\sigma^2)+A \nonumber \\
      &   & +B_{\varphi} \frac{y}{|x|} \left[ 1 - 
    \left(\frac{y}{x}\right)^2
    \right]^{-1/2} \nonumber \\
      &   & \frac{L_2-(x^2+y^2)^{1/2}}{L_2-L_1} )
  \frac{A}{L_2} \frac{1}{2} (x^2+y^2)^{-1/2} 2x
\end{eqnarray}

And for y=0 (warp plane)
\begin{equation}
  \label{fiwp}
  F_z=\frac{A^2}{4\pi L_2} \left( 1 - e^{-\frac{x^2}{2 \sigma^2}} \right)
\end{equation}

The equation of vertical equilibrium becomes
\begin{equation}
  \label{eve}
  \frac{\partial p}{\partial z}+\rho GMr^{-3}z=\frac{B_{\infty}}{8 \pi
  L_2} \left( 1 - e^{-\frac{x^2}{2 \sigma^2}} \right)
\end{equation}
To calculate the warp curve, we use $\frac{\partial p}{\partial z}=0$
as before. We now find for $z$ small (so that $r \approx x)$
\begin{equation}
  \label{zmod2}
  z=\frac{B_{\infty}}{8 \pi L_2 G M \rho_0} e^{\frac{x}{r}}  \left( 1 - e^{-\frac{x^2}{2 \sigma^2}} \right)
\end{equation}
With the same assumptions the first model would give
\begin{equation}
  \label{zmod1}
  z=\frac{B_{\infty}}{8 \pi L_2 G M \rho_0} e^{\frac{x}{r}} 
\end{equation}
We note, however, that the last exponential function in
eq. (\ref{zmod2}) is not very important. For $x=\sigma$, we have
$e^{-\frac{x^2}{2 \sigma^2}}=0.61$; for $x=2\sigma$, we have 0.14; for
$x=3\sigma$, we have only 0.01. Therefore, we conclude that the use of
the first model is fully justified.

\section{Conclusions}

In order to study the influence of intergalactic magnetic fields on
the warping of galactic discs, it is necessary to know their
distribution in the outermost disc, $\vec{B}(x,y,z)$. At present, no
observations are available to determine the field at very large
galactocentric radii. However, we have proposed a field distribution
which should not differ very much from the real one, as the boundary
conditions and the condition $\nabla \cdot \vec{B}$ are in practice
very restrictive. Magnetic field lines must be very similar to those
presented in our figures \ref{fig3}, \ref{fig4} and \ref{fig5}. 

We have shown that the detailed knowledge of the magnetic field lines
in the periphery of the galaxy is relatively unnecessary to predict
reasonable warp curves. More sophisticated calculations should be
required in the future, but considering the present state of our
understanding of warps, a simplified model has more advantages. The
magnetic model of warps provides curves in good agreement with real
warps. Under the magnetic hypothesis, the fitting of the warp geometry
could provide information about two important cosmological parameters:
the direction and the strength of the large scale extragalactic
magnetic field.


\begin{thebibliography}{}

\bibitem[1990]{battaner}
Battaner, E., Florido, E., S\'anchez-Saavedra, M. L. 1990, A\&A, 236, 1
\bibitem[1991]{binney}
Binney, J. 1991 in {\em Dynamics of Disk Galaxies}, ed. B. Sundelius. G\"oteborg.
Sweden
\bibitem[1993]{cuddeford}
Cuddeford, P., Binney, J. J. 1993, Nature, 365, 20
\bibitem[1994]{combes}
Combes, F. 1994 in {\em The Formation and Evolution of Galaxies}, ed. C.
Mu\~noz-Tu\~n\'on \& F. S\'anchez, p. 317. Cambridge University Press.
\bibitem[1995]{dubinski}
Dubinski, J., Kuijken, K. 1995, ApJ, 442, 492
\bibitem[1995]{feretti}
Feretti, L., Dallacasa, Giovannini, G., Tagliani A. 1995, A\&A 302, 680
\bibitem[1959]{kahn}
Kahn, F. D., Woltjer, L. 1959, ApJ, 130, 705
\bibitem[1994]{kronberg}
Krongerg, P. P. 1994, Rep. Prog. in Phys., 57:4, 325
\bibitem[1997]{kuijken}
Kuijken, K. 1997 in {\em Dark Dark and 
Visible Matter in Galaxies and Cosmological Implications} eds. M. Persic \&
P. Salucci. A. S. P. Conf. Series
\bibitem[1988]{sparke}
Sparke, L., Cassertano, S. 1988 MNRAS 248, 58
\bibitem[1997]{zurita}
Zurita, A., Battaner, E. 1997, A\&A, to be published
\end{thebibliography}
\end{document}